# Synthesis of Adversarial DDOS Attacks Using Tabular Generative Adversarial Networks

Abdelmageed Ahmed Hassan
Faculty of Engineering
University of Ottawa
Ottawa, Canada
ahass202@uottawa.ca

Mohamed Sayed Hussein
Faculty of Engineering
University of Ottawa
Ottawa, Canada
mhuss073@uottawa.ca

Ahmed Shehata AboMoustafa
Faculty of Engineering
University of Ottawa
Ottawa, Canada
aabom018@uottawa.ca

Sarah Hossam Elmowafy
Faculty of Engineering
University of Ottawa
Ottawa, Canada
selmo028@uottawa.ca

*Abstract*—Network Intrusion Detection Systems (NIDS) are tools or software that are widely used to maintain the computer networks and information systems keeping them secure and preventing malicious traffics from penetrating into them, as they flag when somebody is trying to break into the system. Best effort has been set up on these systems, and the results achieved so far are quite satisfying, however, new types of attacks stand out as the technology of attacks keep evolving, one of these attacks are the attacks based on Generative Adversarial Networks (GAN) that can evade machine learning IDS leaving them vulnerable.

This project investigates the impact of the Adversarial Attacks synthesized using real DDos attacks generated using GANs on the IDS. The objective is to discover how will these systems react towards synthesized attacks. marking the vulnerability and weakness points of these systems so we could fix them.

*Keywords—IDS, GANS, ML, Cyber Security, AI*

## 1- Introduction

Cyber Attacks are increasingly sophisticated, hackers keep adapting their strategies to exploit every possible vulnerability that might be found in a system or a network, one of the many types of the criminal activities that occur on the web is Distributed Denial of Service (DDos) Attacks, these attacks have the ability to hamper or completely prevent users from accessing networks or websites no matter how robust or large they are, these attacks flood the network with traffic until the network crashes and the servers don't respond, sometimes it takes several hours to be restored preventing the system from offering regular services to legitimate users. DDos attacks can also penetrate a large number of compromised systems called Botnet and launch coordinated attack on the victim system as shown in figure 1. According to Cloudflare, just over one in five DDoS attacks was accompanied by a ransom note from the attacker during 2021. In December, a prime time for online retailers in the run up to Christmas, one in three of the organizations surveyed said they received a ransom letter relating to a DDoS attack [1]. A lot of organizations have been facing potential threats to their network environment that may lead to serious impacts on their operations including data breach, business downtime, unavailability and even ransom demands from hackers. Upon the occurrence of these attacks, actions are needed to be taken.

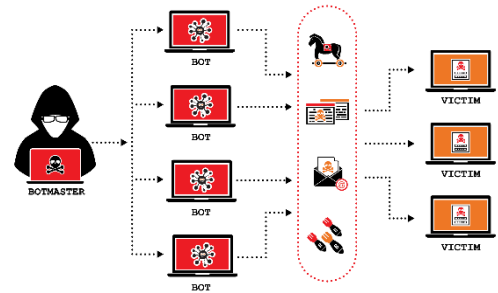

Fig. 1. DDos Attack.

Intrusion Detection Systems have exploited the great potential in AI (Artificial Intelligence) and ML (Machine Learning) and applied them in DDos detection, many achievements have been done by ML in DDos detections as reported in [2], by using ML and AI, IDS can improve their performance by using previous data and predict the attack for the future. With Unsupervised Machine Learning, IDS systems can predict the attacks that aren't labeled but that techniques are prone to False positives [3], this gives the attackers the chance to mislead models into their desired misclassification by using adversarial examples.

The most common Machine Learning algorithms used for Network Intrusion Detection Systems are Random Forests, Decision Trees, Logistic Regression and Naïve Bayes.

Random Forests classifier gives leading performance in designing IDS that are effective and efficient for intrusion detection. Decision Trees is one of the most powerful systems that can handle the intrusions of the computer environments by triggering alerts to make the analysts take actions to stop this intrusion as mentioned here [4]. Applying Naïve Bayes on reduced datasets for intrusion detection gives better performance to design effective and efficient IDS [5]. Also, Logistic Regression (LR) with features selection methods achieves good results with IDS [6].

Generative Adversarial Networks (GANs) are the common method for adversarial attacks. GANs are one of the most important techniques that are used in many areas of Computer Science, it's basically composed of two functional blocks, namely generator and discriminator that play a min-max game to converge to an optimal solution. The generator learns to generate data that looks like the original data or authentic data while the discriminator learns to differentiate that generated fake data from the real original data, the purpose is for the generator to fool the discriminator and generate data indistinguishable from the real one. GANs achieved high performance with images [7], acoustics [8] and textual data [9].

In this project we will use GAN to generate malicious attacks and see if the IDS systems will be able to detect them or not, if they penetrate the IDS systems then this will be a pessimistic call to find countermeasures to solve this problem, also, these fake data may be used as a new signature for IDS systems. GANs maybe a good approach to help IDS system to be more robust in attacks detection.

## 2- RELATED WORK

Machine Learning techniques have been already used in IDS, mainly as classifiers, Y. Li [10] Use a combination of Long Short-Term Memory (LSTM) and Bayesian algorithms to detect DDoS attacks. LSTM can process a long sequence of data with long intervals and delays in time, as it can capture the long-term temporal dependencies in the coming data. So, its neuron output indicates whether to pass the data or neglect it. The author here uses Bayesian method as a second stage to improve the detection performance. Also, Square GAN (LSGAN) [11] used to generate artificial traffic. Almost 99% of DDoS data generated was classified incorrectly as legitimate by the IDS. There is another successful achievement by W. Hu and Y. Tan [12] MalGAN used to generate adversarial malware dataset, which are able to bypass black-box machine learning algorithm-based detection models. The black box detector acts as the target device is hacked. Martin Arjovsky, Soumith Chintala, Léon Bottou [13]

In our project WGAN will be used as it solved a lot of shortcomings found in the original GANs, it is another new approach from GAN, which improves the learning process, makes the algorithm more stable against different variation of data and get rid of problems like mode collapse. Earlier specifically in 2017, Ravi Chauhan, Shahram Shah Heydari [14] They used the newly WGAN algorithm to generate the adversarial data which able to deceive the black-box IDS, then they feed the generated adversarial data to a bunch of machine learning algorithms as Naive Bayes, Logistic Regression, Decision Tree, and Random Forest. These algorithms produce different accuracies, 84.9, 79.5, 75.13, 94.3 respectively, and in final stage they tested the trained IDS with updated attacks can pass the IDS. K. Masataka and O. Kaoru and D. Mianxing [15], they use GAN model to eschew the malware detection system, they tried some machine learning algorithms as Random Forest and Multi-layer Perceptron which gives a high accuracy with 95.6% and 95.5% respectively.

## 3- METHODS

### 3.1 Data Exploration

In data exploration, our objective This phase allows you to learn about a dataset's properties and possible problems without having to make any assumptions about the data. We may next see the correlation between features by displaying correlation matrix, scatter matrix and feature distributions.

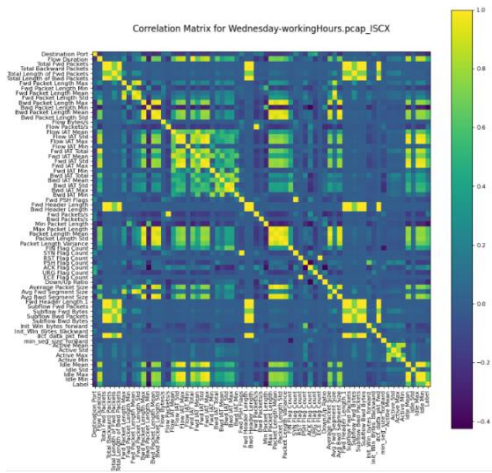

Figure 2. Correlation Matrix of data

As shown in Fig [2], the correlation between the variables on each axis is shown in each square. Correlation might be anywhere between -1 and +1. Closer to 0 indicates that there is no linear relationship between the two variables. The closer the correlation is to +1, the more positively associated they are that is, when one rises, so does the other, and the closer to one they are, the stronger the association. A -1 correlation is similar, except that instead of both variables rising, one will fall when the other rises., one will decrease as the other increases. The diagonals are all 1 since the squares connect each variable to itself (therefore it's a perfect correlation) (yellow). The stronger the connection between the two variables, the greater the number and the lighter the color.

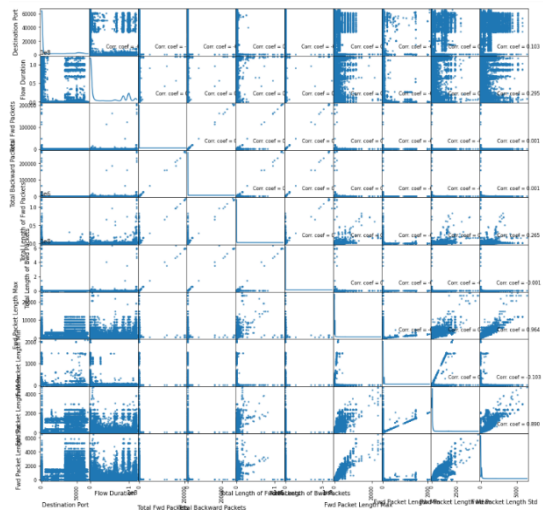

*Figure 3. Scatter and Density plot*

We may generate a quantitative assessment of the strength and direction of the correlation by looking at the scatterplot matrix and getting a visual feel of the variables and their relationships. There are various correlation measurements available. The Pearson product moment (or Pearson's) is one of the more common Fig [3] shows that the matrix's primary diagonal is 1. This indicates that each variable is perfectly correlated with the others. Note that the "-"sign denotes a negative correlation, whereas the lack of a "-"sign implies a positive correlation elsewhere in the matrix. Pearson's values vary from -1.0 (completely negatively correlated) to +1.0 (totally positively correlated). A stronger linear relationship is indicated by values closer to -1.0 or +1.0.

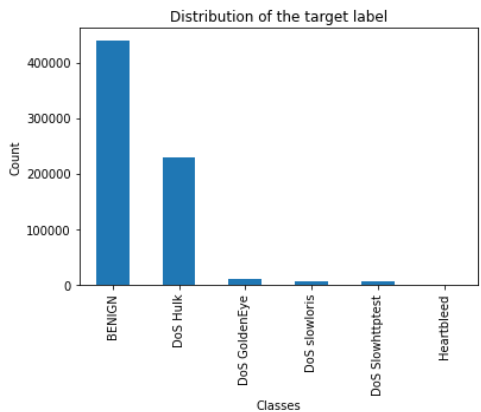

Figure 4: Distribution of target labels

The distribution of each category is BENIGN = 40031, DOS Hulk = 231073, DoS GoldenEye = 10293, DoS Slowloris = 5796, DoS Slowhttptest = 5499 and Heartbleed = 11.

### 3.2 Data Preprocessing

Data is almost good except there are small number of missing values and infinities, so We handled missing values and infinities by removing them because they are a small number and do not affect the distribution of our dataset. Our target label has six categories BENIGN and different types of DOS attack (DoS Hulk, DoS GoldenEye, DoS slowloris, DoS Slowhttptest and Heartbleed). We will handle this by encoding all DOS attack (DoS Hulk, DoS GoldenEye, DoS slowloris, DoS Slowhttptest and Heartbleed) to 1 and BENIGN to 0 and solve it as a binary classification problem.

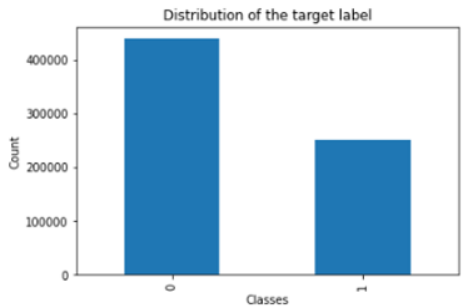

Figure 5: Distribution of the target labels

0 for benign and 1 for malicious, and as shown in Figure 4, we have 440031 rows for benign and 252672 rows for malicious, thus our data is nearly balanced, and no oversampling approaches are required.

### 3.3 Feature Engineering

Here we will use SHAP library [16] to get the most important features, the concept underlying SHAP feature importance is straightforward: It's crucial to have features with high absolute Shapley values. We average the absolute Shapley values per feature across the data to get the global importance:

$$I_j = \frac{1}{n} \sum_{i=1}^{n} |\phi_j^{(i)}|$$

The features are then sorted and plotted in order of decreasing relevance. The importance of SHAP features for the Random Forest trained before for predicting malicious data is shown in the following graphic.

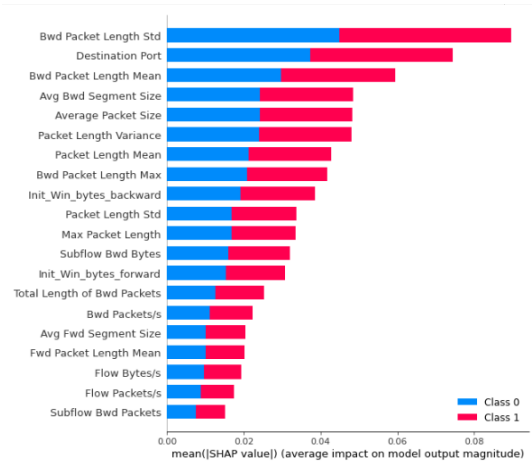

Figure 6: Most Important Features

so, we select the first 10 features to train our models and select the best one to proceed with it.

### 3.4 Models

3.4.1 Random Forest:

Random Forest is a classification technique that uses numerous decision trees to classify data. When constructing each individual tree, it employs bagging and feature randomization to generate an uncorrelated forest of trees whose committee forecast is more accurate than any one tree's [17].

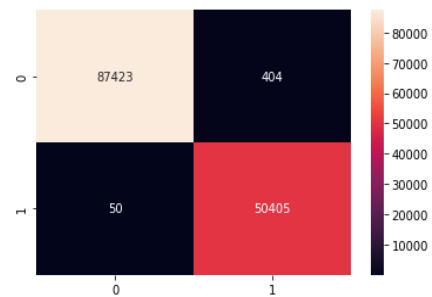

### 3.4.2 Decision Tree

Decision tree is the most powerful and popular tool for classification and prediction. A Decision tree is a flowchart like tree structure, where each internal node denotes a test on an attribute, each branch represents an outcome of the test, and each leaf node (terminal node) holds a class label [18].

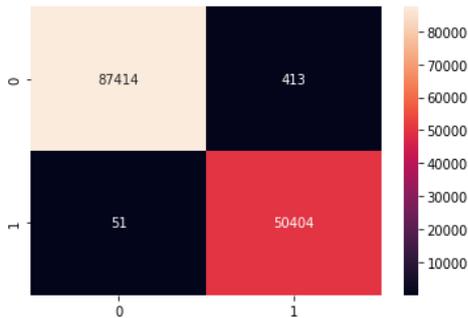

### 3.4.3 Logistic Regression

In its most basic form, logistic regression is a statistical model that utilizes a logistic function to represent a binary dependent variable [18].

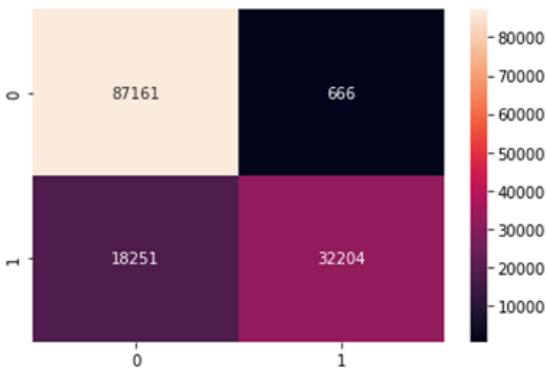

### 3.4.4 Naïve Bayes

The Bayes Theorem is used to create a statistical categorization approach known as Naive Bayes. It's one of the most basic supervised learning algorithms available. The Naive Bayes classifier is a quick, accurate, and trustworthy method. On big datasets, Naive Bayes classifiers exhibit good accuracy and speed [19].

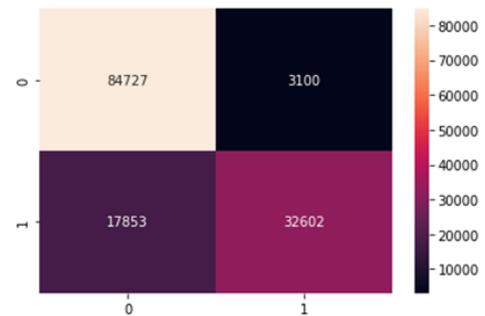

As we can see from the figures the accuracy of Random Forest and Decision tree is almost the same. So, since our problem is sensitive to detecting DOS attacks, we need to look at different metrics to choose the optimal one. We should concentrate on False Negative and choose the minimum value. From

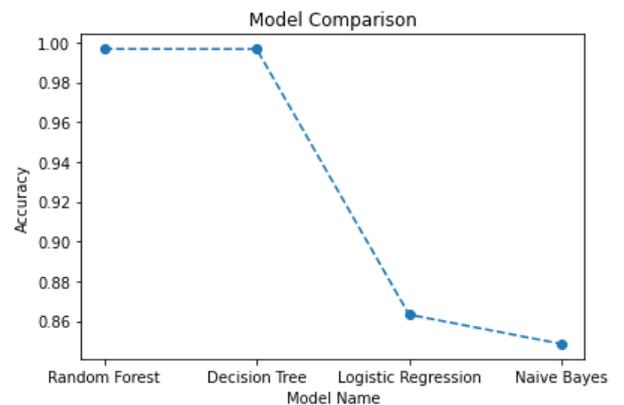

Figure 7: Model Comparison based on Accuracy

## 3.5 Generating Data with Tabular GANs

The main goal of GANs is to generate fake data that is similar to the original data. GANs are made up of two Neural networks that compete: generator and discriminator. The generator tries to generate new samples from noise (usually Gaussian distribution) in an attempt to mimic the original data to deceive the discriminator model Fig [8], which tries to figure out if the data given by the generator is real or fake [21].

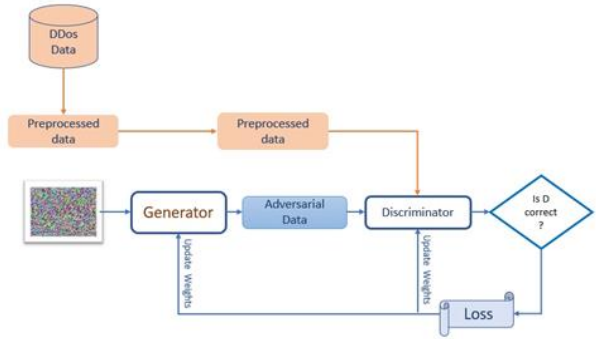

Figure 8: Generative Adversarial Networks GAN

### 3.5.1 Training Table Preprocessing

Preprocessing numerical variables using Gaussian Mixture Model (GMM) as neural networks can effectively generate values with a distribution centered over (-1, 1) as neural networks fail to generate suitable data with multi-modal data] preprocessing categorical variables as SoftMax can be used to directly generate the probability distribution before passing malicious data to the generator. Categorical variables, on the other hand, must be converted to binary variables by one hot-encoding with noise [22]. So each numerical variable is created in two steps, using GMM to create normalized continuous variables and their probability, and each categorical variable is created in one step using one-hot encoding.

### 3.5.2 Generative Model:

The model takes malicious data and attempts to generate a fake one with a similar distribution, using the LSTM with attention mechanism to generate the desired row [23], with numerical variables generated by applying tanh to the cluster vector and categorical features generated as a probability distribution over all possible labels using softmax Fig [9].

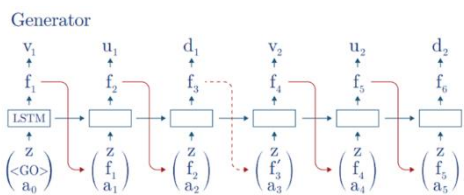

Figure 9: Tabular GANs Generator with LSTM [23]

### 3.5.3 Discriminative model:

The Discriminator's job is to figure out if the data given by the generator is real or fake. To correctly identify real-world data, the discriminator is first trained on it. The Discriminator will be unable to effectively recognize bogus images if it is not adequately taught, resulting in poor Generator training. To distinguish between actual and bogus data, the discriminator concatenates all features and uses Multi-Layer Perceptron [23]. Tables are used to implement GANs (rows and columns of categorical and numerical data) Fig [10].

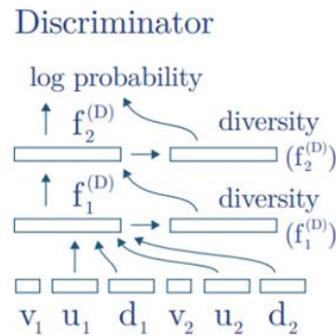

Figure 10: Tabular GANs Discriminator with MLP [23]

## 4- EXPERMENTAL SETUP AND RESULTS

### 4.1 Kafka setup and streaming processing

Apache Kafka is a distributed data storage designed for real-time ingest and processing of streaming data.[24][25]. We used Kafka for sending real time streaming data for our machine learning model to make real time predictions about the coming traffic in order to take the right decision at the right time Fig [11]

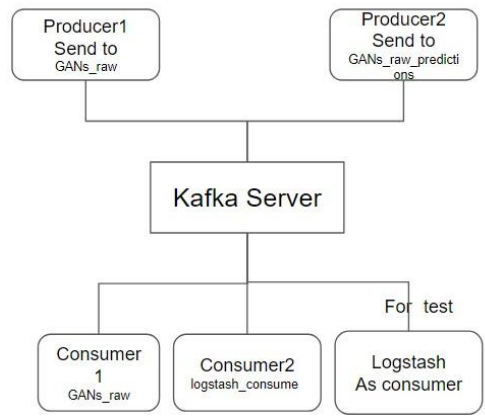

Figure 11: Our Kafka system design

We have made two Kafka producers. Send data to the "GANs raw" topic, and malicious attacks to the GANs raw predictions topic. Then Kafka consumer is used to read data from "GANs raw" topic in order to make predictions, then the malicious traffic to elastic search to be visualized on Kibana

**4.2 Attack visualization environment - ELK Stack**

Elasticsearch, Logstash, and Kibana [26] make up the ELK stack, which is an acronym for a stack made up of three notable projects. The ELK stack, also called as Elasticsearch, provides the possibility of collection logs from most of the systems and applications, analyze them, and visualize them for application and infrastructure monitoring, speedier troubleshooting, security analytics Fig [12].

- E refers to Elasticsearch, which is used as a storage system for logs.
- Logstash is a term that refers to a Logstash that is used for both sending and processing and storing logs.
- K stands for Kibana, which is a web-based visualization tool hosted by Nginx or Apache

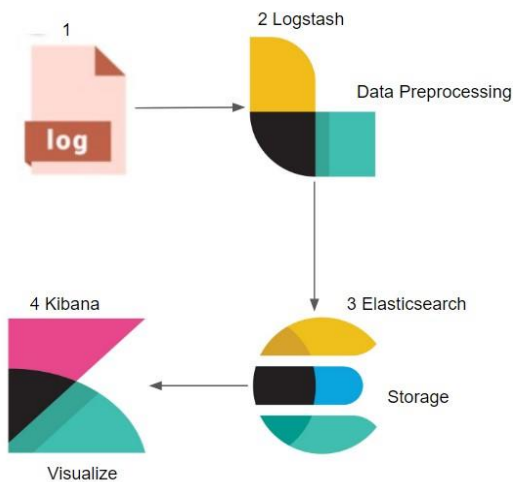

Figure 12: ELK Stack

```
input {
  kafka {
    codec => json
    bootstrap_servers => "localhost:9092"
    topics => ["GANs_raw_predictions"]
  }
}

output {
  elasticsearch {
    hosts => ["localhost:9200"]
    index => "attack_index"
  }
}
```

Server logs, proxy logs, and other logs may be used.

The logs that need to be examined have been identified.

Logstash:

- Gather data from logs and events.
- It can also parse and transform data.

Elasticsearch stores, searches, and indexes the modified data from Logstash.

Kibana, Elasticsearch DB is used by Kibana to Explore, Visualize, and Share data.

**4.2.1 Elasticsearch**

Elasticsearch is an Apache Lucene-based distributed search and analytics engine. Elasticsearch is a good alternative for many log analytics and search use cases due to its support for several languages, high performance, and schema-free JSON data. [27]

**4.2.2 Logstash**

The data collecting pipeline tool is Logstash. It gathers data inputs and sends them to Elasticsearch for indexing. It collects data from a variety of sources and makes it available for future use. Logstash can bring together data from a variety of sources and normalize it for use in your intended destinations. It enables you to cleanse and democratise all of your data in preparation for analytics and use case visualization. It is made up of three parts: **Input:** sending logs to be processed into a machine-readable format, for example, the Kafka topic. **Filters**: A set of requirements to perform a specific action or event **Output**: Decision maker for processed event or log as Elasticsearch

**4.2.3 Kibana**

The ELK stack is completed by Kibana, a data visualization tool. This tool is used to visualize Elasticsearch documents and provides developers with a rapid overview. To make visualization for complex queries, we used the Kibana dashboard which provides a bunch of interactive diagrams, and graphs that gives a good representation for data.

It's a tool for searching, viewing, and interacting with data in Elasticsearch directories. Kibana is a data visualizations tool that allows you to perform extensive data analysis and represent your data in a variety of tables, charts, and maps.

Kibana offers the following features: Visualization is simple Elasticsearch is fully integrated. Simple and intuitive user interface. Capabilities for real-time analysis, graphing, summarization, and debugging.

**4.2.4 Steps**

First, we make Logstash configuration as following. Here, we specify the input source which is Kafka which running on port 9092 and consume data from Kafka topic GANs_raw_predictions , then, the output is Elasticsearch running on port 9200 and index attack index to be used on Kibana to receive the data

Figure 13: Logstash Configuration

Here, we specify the input source which is Kafka which running on port 9092 and consume data from Kafka topic GANs_raw_predictions, Then, the output is Elasticsearch

running on port 9200 and index attack index to be used on Kibana to receive the data

### 4.2.5 Dashboard and results

We configure the Elasticsearch port to 9200 and Kibana to port 5601 , after running our Kafka server and model predict on our data generated by GANs, we restrict the label generated to attack only to be sent to Logstash, after that, we run the Kibana and create our real-time dashboard as following

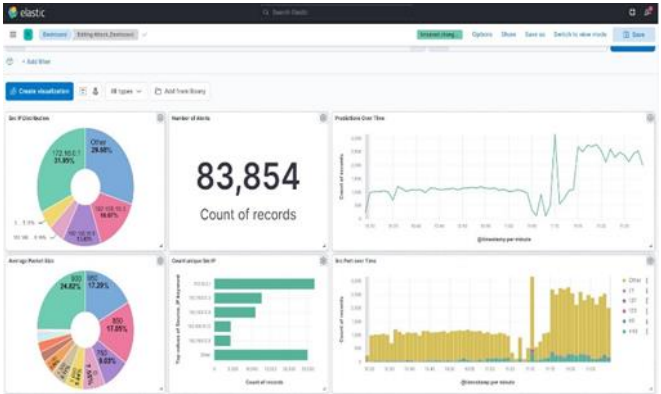

Figure 14: Kibana Dashboard

The count of records is 83,854 records and is changes over time, In the pie chart on the top left, we see the Source IP of attack devices, the pie chart below it, it represents the average packet size, on the top right we see the line chart represents the predictions over time.

## 5 - CONCLUSION AND FUTURE WORK

Duo to the rapid growth of Internet usage and the generation of huge amounts of valuable digital data, the attackers have been attracted and tempted to illegally attain benefits. This paper introduced one way of limiting such illegal actions by synthesizing new types of attacks using GANs and testing them using four different Intrusion Detection System. Also, we have done some experiments on CICIDS.2017 up to date data set using the same algorithms and the results are compared. The models achieved high results in terms of Accuracy and Recall on both datasets. Although the accuracy does not depend only on the algorithm but also on the application area, so in the future we will conduct an exclusive study of ML algorithms to provide better solutions for the IDS on real time datasets.

## 6- REFRENCES

[26] ELK Stack Tutorial: What is Kibana, Logstash & Elasticsearch? (n.d.). https://www.guru99.com/elk-stack-tutorial.html. https://www.guru99.com/elk-stack-tutorial.html

[27] Free and Open Search: The Creators of Elasticsearch, ELK & Kibana Elastic. (n.d.). https://www.elastic.co/. https://www.elastic.co/